\documentclass[%
iop,
amsmath,amssymb,
reprint,%
]{revtex4-1}
\usepackage[utf8]{inputenc}
\usepackage[english]{babel}
\usepackage{amssymb}
\usepackage{amsmath}
\usepackage[pdftex]{graphicx}
\usepackage{cmap}
\usepackage[pdftex]{hyperref}
\usepackage{float}

 \def\bc{\begin{center}}          \def\ec{\end{center}}
 \allowdisplaybreaks

\begin{document}

\title{Electromagnetic emission due to nonlinear interaction of laser wakefields colliding in plasma at an oblique angle}
 \author{E. Volchok}
 \affiliation{Budker Institute of Nuclear Physics SB RAS, 630090, Novosibirsk, Russia}
 \affiliation{Novosibirsk State University, 630090, Novosibirsk, Russia}
 \author{V. Annenkov}
 \affiliation{Budker Institute of Nuclear Physics SB RAS, 630090, Novosibirsk, Russia}
 \affiliation{Novosibirsk State University, 630090, Novosibirsk, Russia}
 \author{I. Timofeev}
 \affiliation{Budker Institute of Nuclear Physics SB RAS, 630090, Novosibirsk, Russia}

 \begin{abstract}
 	
	Head-on collision of laser-induced plasma wakefields with differing profiles of electrostatic potential has been recently found to be an efficient mechanism for generating high-power electromagnetic emission at the second harmonic of the plasma frequency [I.V.Timofeev et al. Phys. Plasmas {\bf 24}, 103106 (2017)]. This mechanism is attractive for creating a source of tunable narrow-band coherent radiation in the terahertz frequency range. In this paper, we generalize the theory of electromagnetic emission produced by nonlinear interaction of two plasma wakes to the case of an arbitrary collision angle. Such a theory is used to evaluate the angular distribution of the second harmonic radiation as well as its total generation efficiency for parameters of the proof-of-principle experiment in which laser axes will be aligned with a small finite angle. Theoretical predictions are qualitatively confirmed by particle-in-cell simulations.

 \end{abstract}
 \maketitle

\section{Introduction}

Investigation of the mechanisms responsible for electromagnetic (EM) waves generation at the plasma frequency harmonics refers to one of the most fundamental problems of plasma physics. On the one hand, this problem arises in the context of solar radio bursts physics \cite{Ginzburg1958, Gurnett1976, reid2014review, krafft2019electromagnetic}. On the other hand, such emission serves as an indicator of the intense Langmuir turbulence in laboratory beam-plasma experiments  aimed at developing techniques for plasma heating and confinement in open magnetic traps \cite{burdakov2011concept,arzhannikov2013experimental,Burdakov2013, annenkov2019highly}. In addition to the fundamental component of this issue, emissions at harmonics of the plasma frequency are also of interest with regard to generation of  high-power THz radiation. Such radiation is in demand for many applications in the range from security systems to material science \cite{federici2005thz,abbasi2016nano, tonouchi2007cutting}. It is of particular interest to study the properties of materials and methods of subtle impact on the states of matter. EM radiation in the terahertz range provides resonant and nonresonant  access to the fundamental modes of matter such as crystal lattice vibrations, molecule rotations and precession of spins \cite{cole2001coherent, kampfrath2013resonant, hafez2016intense}.

Despite the relevance of the topic, the problem of the so-called {\it terahertz gap}  exists to this day. It consists in the absence of powerful and compact radiation sources in this frequency range. Although the tremendous progress has been achieved in producing powerful single-cycle THz pulses \cite{wu2013intense, Liao3994, Wu2018, Herzer_2018}, generation of longer emission with a narrow line-width and high values of electric field is still a challenging task. One of the ways for this problem to be solved is free electron lasers which are still the most powerful sources of narrow-band radiation in the THz range \cite{vinokurov2011free, felix}. However, they are large and expensive devices, so the question of creating more compact sources  has been opened so far.  Possible solutions are proposed by both acceleration \cite{tian2017femtosecond,li2016generation} and laser \cite{Ahr17, liu2017generation} communities.  It is worth noting the special role of plasma-based schemes in solving this problem. The attractiveness of plasma in terms of THz generation is primarily associated  with the possibility to excite in such a nonlinear medium  long-lived waves with amplitudes exceeding the destruction threshold of matter. The simplicity of tuning the radiation frequency by changing the plasma density makes it possible to cover most of the terahertz band. For efficient excitation of large-amplitude plasma waves, either short laser pulses or relatively long electron beams can be applied. Further transformation of these waves into EM oscillations at the plasma frequency can be realized via the linear mode conversion assuming the presence of density gradients \cite{annenkov2016generation, miao2017high, Arzhannikov2020} or an external magnetic field \cite{wang2015tunable, yugami2002experimental, kwon2018high}. On the other hand, plasma waves can participate in various nonlinear processes generating radiations at higher harmonics. Such radiations are not screened by plasma and can leave it freely. 
The nonlinear emission process which is frequently discussed in relation to the electron beam-plasma interaction is the coalescence of two plane Langmuir waves travelling in opposite directions at an acute angle with respect to each other.  In this case, a counterpropagating wave can arise as a result of filling the spectrum of oscillations during the development of Langmuir turbulence as assumed in solar physics \cite{malaspina2012antenna, henri2019electromagnetic, whelan1981electromagnetic}. This backward wave can also be excited directly by the counter-injection of an additional electron beam, which is proved to enhance the second harmonic emission in both laboratory \cite{leung1981observation, schumacher1993microwave} and numerical \cite{ganse2014fundamental, Timofeev2014} experiments. Recently, it has been found that even head-on collision of electrostatic waves in a uniform plasma can result in the second harmonic EM emission if colliding waves have mismatching transverse profiles of electrostatic potential \cite{timofeev2017generation}. The most simple way to excite Langmuir waves with controlable transverse structures is to irradiate plasma by short laser pulses. The scheme based on using counterpropagating laser drivers produced by PW-class laser systems allows one to generate  terahertz radiation with the gigawatt power level, narrow line-width of $\sim1\%$ and energy conversion efficiency exceeding $10^{-4}$.

A distinctive feature of the proposed mechanism is that the role of drivers of colliding plasma waves can be played not only by short laser pulses, but also by long-pulse electron beams unstable against the two-stream instability \cite{annenkov2018high}. Both generating schemes have their advantages and disadvantages. An unstable electron beam is able to transfer more than half of its initial energy into plasma oscillations, which gives a gain in efficiency compared to the laser scheme. Applying long-duration beams allows one to maintain the amplitudes of generated waves at a high nonlinear level for a longer time. However, the use of electron beams is associated with a number of technical difficulties. Parameters of the beam-plasma instability are more difficult to control and the system itself turns out to be cumbersome and difficult to operate. Contrary, laser pulses are able to transfer only a small part ($\sim 1\%$) of their energy into long-lived plasma modes,  and the duration of the produced radiation is strongly limited by the bulk of energy accumulated initially in plasma wakes. At  the same time, amplitude profiles of laser-induced wakefields are much more controlable, that is why it is easier to realize   the proposed mechanism in laboratory experiment. So, to demonstrate  the performance of this idea experimentally,  it has been decided to use the laser-based scheme. We propose to verify this idea using the multi-TW laser system developed in the Insitute of Laser Physics SB RAS (Novosibirsk) \cite{bagayev2014coherent}. Such a proof-of-principle experiment will allow us to understand whether all fundamentally important effects are taken into account by the developed theory and whether the proposed generation scheme is suitable for scaling to more powerful laser systems. In this experiment, we plan to inject axially symmetric Gaussian laser pulses with $\lambda_0=830$ nm into a supersonic gas jet. Laser pulses are focused in the same spatial point, but in different-size spots, thereby exciting transversely localized wakefields with differing potential profiles.  In order to avoid the return of laser radiation into the amplifying system, it is recommended to align laser axes with a small angle. Thus, the purpose of the present work is to find how the radiation efficiency and its angular distribution will change if we introduce a finite angle in the baseline scenario of the head-on collision described in our previous work \cite{timofeev2020simulations}. It seems obvious that this efficiency will decrease with the growth of angle due to the decrease in the volume of wakes overlapping. However, it is necessary to know exactly how strong this decrease will be for angles required for the experimental implementation of the scheme.
 Since the produced radiation is predicted to be narrowly directed, it is also important to know its angular distribution in order to be able to detect it. This paper presents a generalized analytical theory that describes the nonlinear interaction of plasma wakes propagating at an arbitrary angle to each other. 

 This article is structured as follows. Section \ref{Sec:Theory} is devoted to the analytical theory for the case of  laser wakefields collision at an arbitrary angle. In Section \ref{Sec:Headon}, we study the limiting case of head-on collision and how calculations in the far zone agree with the known solution of the boundary problem.   Section \ref{Sec:ThDiscus} contains an analysis of how the radiation characteristics are modified as the angle changes. In Section \ref{Sec:Simulations}, numerical simulation of this problem in plane geometry is considered.  A summary of the results obtained is presented in Section \ref{Sec:Conclusion}.
 
 \section{Theory \label{Sec:Theory}}

\begin{figure}[hbt]
	\begin{center}
		\includegraphics[width=0.9\linewidth]{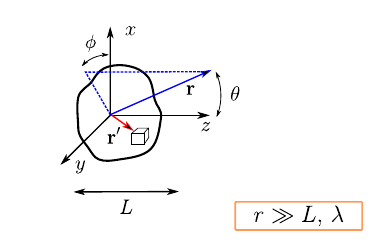}
		\caption{Schematic illustration for the far zone approximation: the distance to the observation point is much greater than both its linear dimensions  $r\gg L$ and the radiated wavelength $r\gg\lambda$. The angles of the spherical coordinate system are introduced as shown.}
		\label{fig:Sytem1}
	\end{center}
\end{figure}

Consider the EM radiation generated by interaction of two plasma waves with different transverse profiles colliding at a certain angle. Let each wave have a frequency $\omega$, then nonlinear interaction of these waves drives oscillations at the doubled frequency $2\omega$. Such oscillations can convert into vacuum EM emission at the same frequency. Misalignment of waves breaks the axial symmetry of the system. In addition, the direct solution of the boundary value problem cannot be used here  as was done in the work \cite{timofeev2017generation}, since such an approach, for simplicity, implied the radiation wavelength to be much smaller compared to the size of the longitudinal spatial inhomogeneity. In the case of oblique collision, this assumption can be applied only at sufficiently small angles. The general case requires another approach. In particular, the characteristics of the radiation can be obtained in the far zone of the source. In the approach utilized in this paper, we neglect the influence of plasma boundaries considering the source placed in an infinite plasma. Also we assume source dimensions spatially limited  and the distance to the observation point is much greater than its all linear dimensions  $r\gg L$ and the radiated wavelength $r\gg\lambda$ (Fig.\ref{fig:Sytem1}). At the same time, we will continue to be interested in the spatial structure of waves in the overlapping region. So, although the dimensions of the source are small, it is not pointlike on the scale of the radiation wavelength.

\subsection{Emission in the far zone}

First, let us focus on general properties of radiation at a large distance from the source without considering the waves structure and features of their origin. For this purpose, consider the expression for the vector potential of the magnetic field generated by electric current oscillations
\begin{equation}
\textbf{A}(\textbf{r}, t)=\frac{1}{c} \int \frac{\textbf{j}(\textbf{r}^{\prime},t^{\prime})}{|\textbf{r}-\textbf{r}^{\prime}|} d^3r^{\prime}, \label{A}
\end{equation}
where $\textbf{j}$ is the density of the radiating current, ($\textbf{r}^{\prime}$, $t^{\prime}$) and ($\textbf{r}$, $t$)  are coordinates of points inside and outside the source, respectively. The radiating current can be represented in the form $\textbf{j}(\textbf{r}, t)=\textbf{J}({\bf r})e^{-2 i \omega t}+c.c.$. Using the conditions $r\gg L,\lambda$ and accounting for the delay $t^{\prime}{=}t-|\textbf{r}-\textbf{r}^{\prime}|/v_{c}$ ($v_c$ is the speed of light in a medium), we obtain a component of $\textbf{A}$ responsible for the dipole radiation:
\begin{equation}\label{Ad}
\textbf{A}(\textbf{r}, t)=\frac{1}{c r}e^{ikr-i 2 \omega t}\int\textbf{J}(\textbf{r}^{\prime})e^{-i\textbf{k}\textbf{r}^{\prime}}d^3r^{\prime}.
\end{equation} 
In this expression, ${\bf k}=(k\sin\theta\cos\varphi, k\sin\theta\sin\varphi, k\cos\theta)$ denotes the wave vector of the emitted EM wave. The length of this vector is determined by the linear dispersion equation for EM oscillations in a cold plasma,  $k c=2 \omega \sqrt{\varepsilon}$, where $\varepsilon{=}1 {-} \omega_p^2/(2\omega)^2$ is the corresponding dielectric permittivity ($\omega_p{=}\sqrt{4 \pi n_0 e^2/ m_e}$ is the plasma frequency, $n_0$ is the unperturbed plasma density, $e$ and $m_e$ -- the charge and mass of an electron). Being comparable in dimensions with the radiation wavelength, the extended source can be represented by a set of radiating dipoles, which is  taken into account by the phase delay $e^{-i\textbf{k}\textbf{r}^{\prime}}$. 
Since radiation in the far zone  is considered as a plane EM wave propagating from the source, the intensity of radiation  carried by the EM wave into a solid angle element takes the form
\begin{equation}
\frac{dI}{d\Omega}=\frac{k}{4 \pi\omega \varepsilon} \left|\int \left[ \textbf{J} \times \textbf{k}\right] e^{- i \textbf{k}\textbf{r}^{\prime}} d^3r^{\prime} \right|^2. \label{dIdW}
\end{equation}
As in the case of a conventional dipole antenna, the energy flux through an elementary area subtending the solid angle $d\Omega$ is seen not to depend on the distance to the observation point. The expression (\ref{dIdW}) defines the angular distribution of the radiation intensity. This distribution is important for the correct detection of $2\omega_p$-emission in future experiments. 

The total radiation power is found as an integral over the entire solid angle
\begin{equation}\label{P}
P=\int \frac{dI}{d\Omega} \sin\theta d\theta d\phi.
\end{equation}
In dimensionless units it is derived as follows
\begin{equation}\label{Pdim}
\dfrac{P}{P_0}=\dfrac{k}{16 \pi^2  \omega \varepsilon} \int \left|\int \left[ \textbf{J} \times \textbf{k}\right] e^{- i \textbf{k}\textbf{r}^{\prime}} d^3r^{\prime} \right|^2 d\Omega, 
\end{equation}
where the power is measured in units of 
\begin{equation}\label{P0}
P_0=\frac{m_e^2 c^5}{4 \pi e^2}\approx 0.69\ \mbox{GW}.
\end{equation}
Hereinafter, we use the following dimensionless units: electric and magnetic fields are expressed in units of $m_e \omega_p c/e$, potentials -- in $m_e c^2/e$, frequencies and wave vectors -- in units of the plasma frequency $\omega_p$ and $\omega_p/c$, respectively.

\subsection{Radiation source}

\begin{figure}[htb]
	\includegraphics[width=0.9\linewidth]{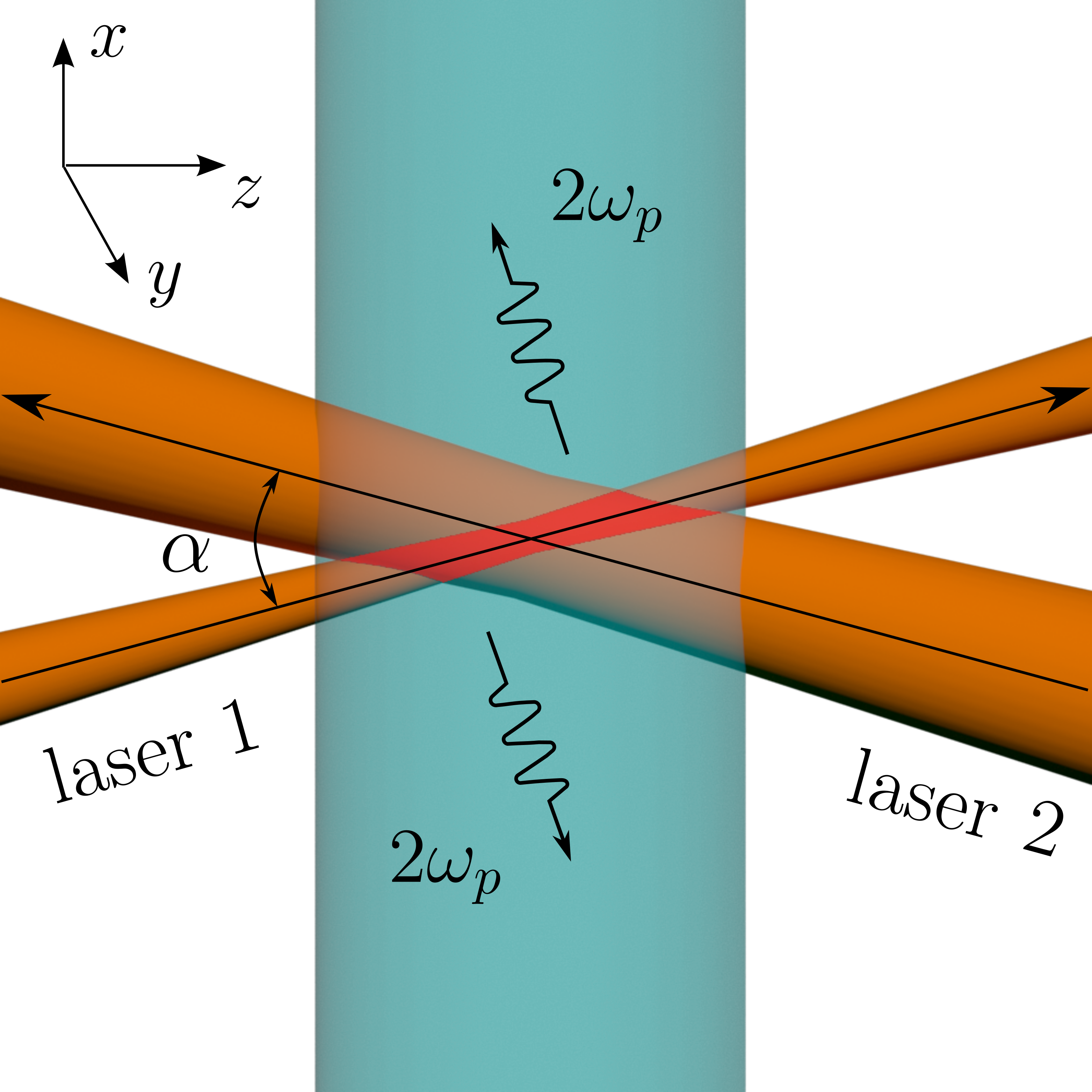}
	\caption{Overlapping scheme of two laser-induced plasma wakes colliding at an angle $\alpha$ in a supersonic gas jet. Radiating region is shown by red.}
	\label{fig:Lasers3D}
\end{figure}

Now let us calculate the current responsible for the second harmonic emission. Consider two laser pulses characterized by the central frequency $\omega_0$ and propagating through the gas at an angle $\alpha$ with respect to each other as shown in Figure \ref{fig:Lasers3D}. Laser radiation ionizes the gas and excites  potential waves with the frequency $\omega=\omega_p$ and longitudinal component of the wavevector $k_{1,2}\approx\omega_p/c$ in the produced plasma. Propagation of each wave is accompanied by harmonic perturbations of the density and velocity of plasma electrons. Mutual scattering of one Langmuir wave on the electron density perturbation produced by a counterpropagating wave results in generation of a nonlinear current capable of emitting EM radiation at the frequency $2\omega$.

Let us write the electric field of excited  wakes in terms of scalar potentials:
\begin{align}
\Phi_{s}(\textbf{r},t)=&\dfrac{1}{2}\left(\Phi_{0s} (\textbf{r}) e^{i \textbf{k}_{s} \textbf{r}- i \omega t}+\mbox{c.c.}\right),\\
&\textbf{E}_{s}=-\nabla \Phi_{s}(\textbf{r},t),
\end{align}
the index $s$ takes values $1,2$, which correspond to the first and second wakes.  Without loss of generality, we assume that the coordinate system is oriented in such a way that the wavevectors lie in the plane  $(x,z)$  and 
\begin{align}
&k_{1x}=  \sin \beta, \quad k_{2x}=  \sin \beta,\\
&k_{1z}=  \cos \beta, \quad k_{2z}= - \cos \beta, \nonumber
\end{align}
where $\beta=\alpha/2$. The radiating current is expressed in terms of density and velocity perturbations of electrons:
\begin{equation}\label{current}
j(\textbf{r},t)=-n^{(1)} \textbf{v}^{(1)}-\textbf{v}^{(2)} =\mathcal{J}(\textbf{r})e^{-2 i \omega t}+c.c.,
\end{equation}
superscripts $(1,2)$ denote the order of perturbation, the current is measured in units of $e n_0 c$. The density and velocity perturbations, written in units of $ n_0 $ and the speed of light $ c $, are derived from the motion equation and the continuity equation and can be represented in the form:
\begin{align}
\textbf{v}^{(1)}_{s}=i \nabla \Phi_{s}&(\textbf{r},t), \quad \textbf{v}^{(2)}_{s}=\dfrac{-i}{2 \omega} (\textbf{v}^{(1)}_{s}\nabla)\textbf{v}^{(1)}_{s}, \nonumber\\
&n^{(1)}_{s}=\dfrac{-i}{\omega}(\nabla\textbf{v}^{(1)}_{s}).
\end{align}

\begin{figure}[h]
	\includegraphics[width=0.99\linewidth]{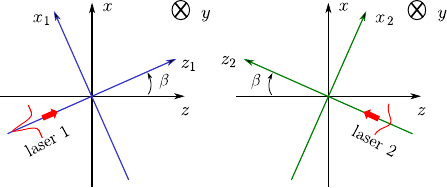}
	\caption{Coordinate systems co-directed to the propagation axes of colliding laser pulses.}
	\label{fig:Coordinates}
\end{figure}

Calculations are simplified if coordinate systems are co-directed to the laser axes as shown in Figure \ref{fig:Coordinates}. The plane $(x, z)$ is rotated around $y$-axis by the angle $\beta$ and $\pi- \beta$.
Corresponding coordinate transformations are
\begin{align}
&x_1=x\cos\beta-z\sin\beta, \qquad x_2=-x\cos\beta-z\sin\beta,\\
&z_1=x\sin\beta+z\cos\beta, \qquad z_2=x\sin\beta-z\cos\beta.\nonumber
\end{align}
In the co-directed frames, EM fields of linearly polarized laser pulses are expressed in terms of the normalized vector potential of the magnetic field
\begin{equation}
\Phi_{s}(\textbf{r}, t)=\int\limits_0^{t} \sin(t-t^{\prime}) \dfrac{a^2_s(\textbf{r}, t^{\prime})}{4},
\end{equation}
which in turn can be represented as follows
\begin{align}
&a_s=a_{0s}\dfrac{\sigma_{0s}}{\sigma_i(z_s)}e^{-(x_s^2+y^2)/\sigma_s(z_s)}\sin^2\left(\dfrac{\pi(t\pm z_s)}{2\tau}\right),\\
&\sigma_s(z_s)=\sigma_{0s}\sqrt{1+z_s^2/\mathcal{R}_s^2}, \qquad \mathcal{R}_s=\omega_0\sigma_{0s}^2/2, \nonumber
\end{align}
where $a_{0s}$ is the maximal value of the laser strength parameter in a focus, $\sigma_{0s}$ is the focal spot-size, $\tau$ is the pulse duration, and $ \mathcal{R}_s$ is the Rayleigh length of each laser beam. As a result, we obtain for the wake amplitudes:
\begin{align}
\Phi_s(&\textbf{r})=\Phi_s^w \left(\frac{\sigma_{0s}}{\sigma_s(z_s)}\right)^2 e^{-2(x_s^2+y^2)/\sigma_s^2(z_s)}, \\
&\Phi_s^w=\frac{3}{4} a_{0s}^2 \dfrac{\sin \tau}{(4-5 \tau^2/\pi^2+\tau^4/\pi^4)}.
\end{align}
We assume that the dependence on the longitudinal coordinate $ \sigma_s (z_s) $ is rather slow and $z_s$-derivatives can be neglected in comparison to derivatives in transverse directions. This condition makes it possible to use the following relations for the derivatives when calculating the contribution of each laser pulse
\begin{align}
&\partial_z=-\sin \beta \: \partial_{x_1}, \quad \partial_z=\sin \beta \: \partial_{x_2}, \\
&\partial_x=\cos \beta \: \partial_{x_1}, \quad \partial_x=\cos \beta \: \partial_{x_2},
\end{align}
where $\partial_z, \partial_x$ and so on are partial derivatives with respect to the corresponding coordinate. Substituting all relevant variables in (\ref{current}) and taking into account the above assumptions, we obtain the electric current responsible for electromagnetic radiation at the second harmonic of the plasma frequency. The resulting expression for this current is rather  lengthy and presented in Appendix \ref{sec:appendix}.

\section{Head-on collision \label{Sec:Headon}}

Let us compare the approach described here with the results obtained in our recent work \cite{timofeev2017generation} for the case of the head-on collision of laser pulses, i.e. at $\alpha=0^{\circ}$. The solution of the boundary problem \cite{timofeev2017generation} predicts that nonlinear interaction of counterpropagating laser wakefields in an axially symmetric plasma column results in emission of EM wave with the amplitude
\begin{align}\label{ampl}
&\mathcal{E}_0= \frac{ 3 \Phi_1^w \Phi_2^w \mathcal{F}_{\sigma}^{b}}{2\sqrt{(2\sqrt{3} R J_1-J_0)^2+16 R^2 J_0^2}}, \\
&\mathcal{F}_{\sigma}^{b}=\frac{\sigma_{01}^2 \sigma_{02}^2 \left|\sigma_2^2-\sigma_1^2\right|}{(\sigma_1^2+\sigma_2^2)^{2}} \exp\left[-\frac{3}{8} \frac{\sigma_1^2 \sigma_2^2}{\sigma_1^2+\sigma_2^2}\right],
\end{align}
where $J_0$ and $J_1$ denote the corresponding Bessel functions of the argument $\sqrt{3}R$, $R$ is the plasma column radius. Amplitudes of wakefields are introduced in the same way as above. The radiation power in this case is given by the integral 
\begin{equation}\label{Pbound}
\frac{P}{P_0}=  \pi R \int\limits_{-\infty}^{\infty} \mathcal{E}_0^2 dz.
\end{equation}

\begin{figure}[htb]
	\includegraphics[width=1.\linewidth]{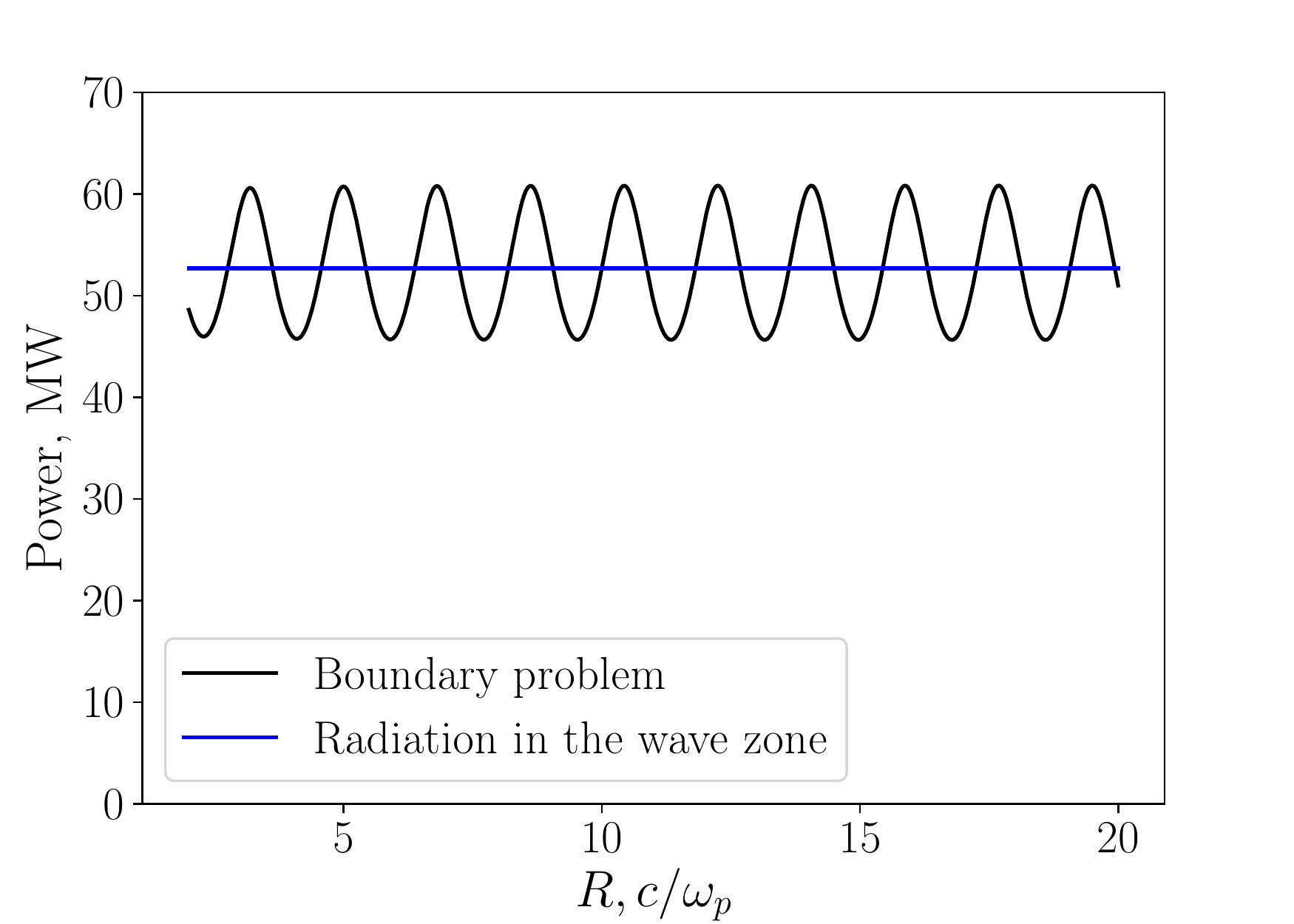}
	\caption{The radiation power as a function of plasma radius obtained in the boundary problem for parameters of the demonstration experiment (black line). The same power calculated in the far zone of the sourse in an infinite plasma (blue line).}
	\label{fig:PowerComp}
\end{figure}

On the other hand, in this particular case, the radiation characteristics can be also calculated in the far zone of the source using the approach of infinite plasma. For $\alpha = 0$, some of the integrals in (\ref{Pdim}) can be approximately calculated analytically. As a result, the radiation magnetic field can be represented in the form
\begin{align}
&B_y=\frac{ik^3}{2r}  \Phi_1^w \Phi_2^w \sin^3\theta e^{ikr-2it}\int F_{\sigma}^{i}e^{-ik z \cos\theta}dz,\\
&F_{\sigma}^{i}=\frac{\sigma_{01}^2\sigma_{02}^2}{16}\frac{\sigma_2^2-\sigma_1^2}{(\sigma_1^2+\sigma_2^2)^2}\exp\left(-\frac{k^2\sin^2\theta \sigma_1^2\sigma_2^2}{8(\sigma_1^2+\sigma_2^2)}\right).
\end{align}
The total radiation power can be written as
\begin{equation}\label{Pwave}
\frac{P}{P_0}=\frac{(\Phi_1^{w}\Phi_2^{w})^2}{2\sqrt{\varepsilon}}\int k^6 \sin^6\theta \left|\int F_{\sigma}^{i}e^{-i k z \cos\theta} dz\right|^2 d\Omega.
\end{equation}
In Figure \ref{fig:PowerComp}, the solution of the boundary problem  for the radiation power (\ref{Pbound}) as a function of plasma radius $R$ (black line) is compared to the same power calculated in the far zone of the source (\ref{Pwave}) (blue line) for parameters of demonstration experiments presented in Table \ref{Table:Params}. As one can see, the presence of the plasma boundary leads to the appearance of oscillations around the mean value that coincides with the power obtained in the far zone. The period of these oscillations is determined by the number of radiation wavelengths fitting into the plasma width. This result allows us to judge the adequacy of the method used and its convergence to the previous results.

\section{Predictions for demonstration experiment \label{Sec:ThDiscus}}

\begin{table*}[htb]
	\centering
	\caption{Parameters of demonstration experiment}
	\begin{ruledtabular}
		\begin{tabular}{lclc}\label{Table:Params}
			
			Laser wavelength, $\lambda_0$	&	$830$ nm	&	Laser pulses duration, $\tau$	&	$39$ fs\\
			
			Plasma density, $n_0$	&	$2.5\cdot10^{18} \mbox{cm}^{-3}$	& 	Frequency of THz radiation, $2\omega_p/(2\pi)$	&	$28.4$ THz \\
			
			Energy of the 1st laser pulse, $\mathcal{W}_{L1}$	&	$16$ mJ	&	Energy of the 2nd laser pulse, $\mathcal{W}_{L2}$	&	$184$ mJ\\
			
			Spot-size of the 1st laser pulse, $\sigma_{01}$	&	$6.3 \mu\mbox{m}$	&	Spot-size of the 2nd laser pulse, $\sigma_{02}$	&	$18 \mu\mbox{m}$\\
			
			Maximal laser strength, $a_{01}$	&	$0.67$	&	Maximal laser strength, $a_{02}$	&	$0.8$\\
			
		\end{tabular}	
	\end{ruledtabular}
\end{table*}

Let us now analyze how the total radiation power and its angular distribution depend on the finite collision angle  for parameters of the planned demonstration experiment which have been discussed in our previous paper \cite{timofeev2020simulations} and are listed in Table \ref{Table:Params}. We consider the interaction of two plasma wakes driven by laser pulses ($\lambda_0=830$ nm) with different energies ($16$ mJ and $184$ mJ) and spot-sizes ($\sigma_{01}=6.3 \mu\mbox{m}$ and $\sigma_{02}=18 \mu\mbox{m}$) propagating in a plasma with the density $2.5\cdot10^{18} \mbox{cm}^{-3}$. In such a plasma, EM radiation is generated at the frequency $2\omega_p/(2\pi)=28.4$ THz. In the case of the head-on collision, these parameters have been found to maximize the radiation efficiency which is defined as the fraction of laser energy converted into the energy of $2 \omega_p$-radiation.

\subsection{Radiation power}

\begin{figure}[htb]
	\includegraphics[width=1.\linewidth]{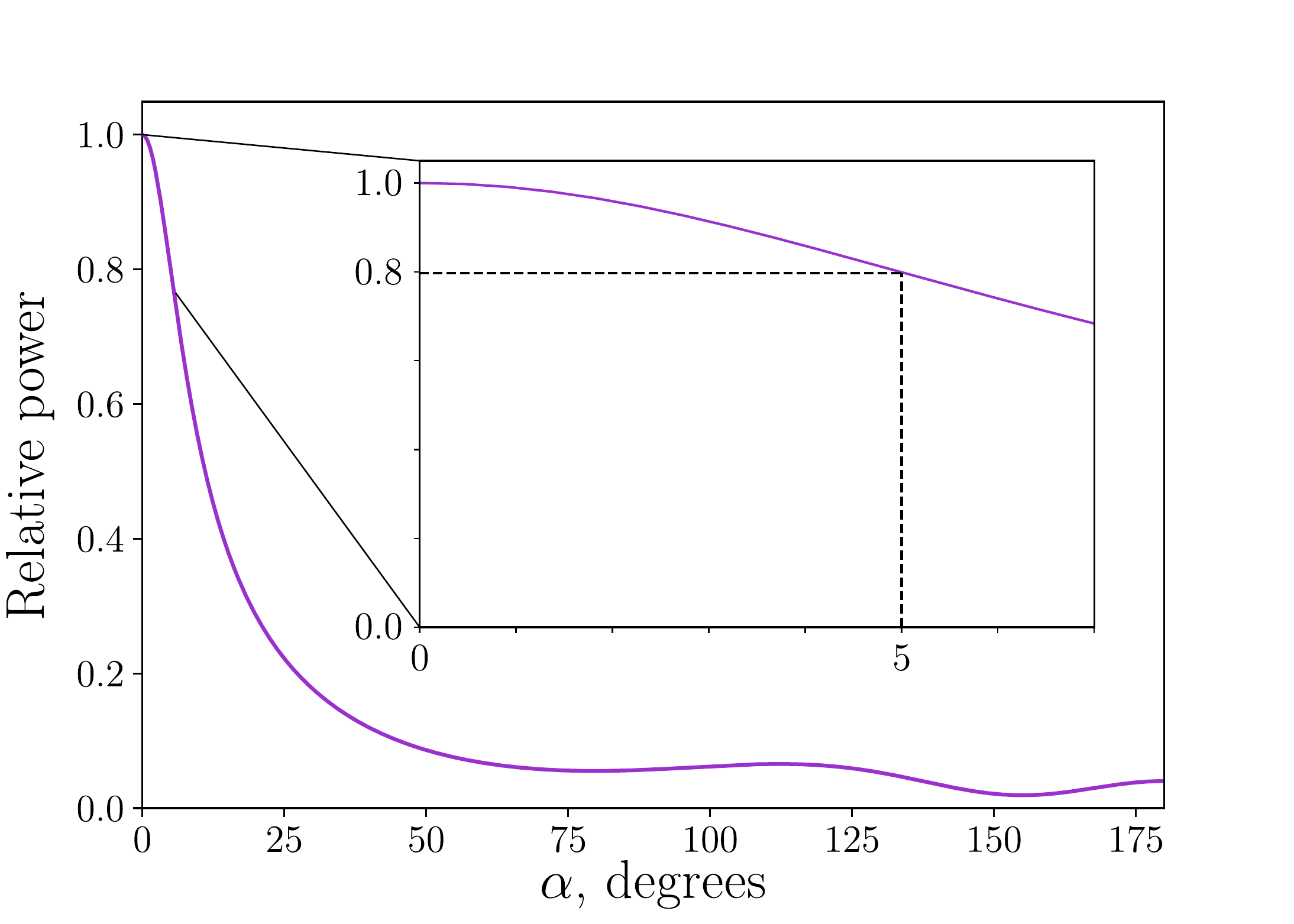}
	\caption{The dependence of the relative power of EM radiation in the far zone (in units of $P(\alpha=0^{\circ})$) on the collision angle $\alpha$.}
	\label{fig:Power}
\end{figure}

Shown in Figure \ref{fig:Power} is  the dependence of the radiation power (\ref{Pwave}) on the collision angle $\alpha$. The power is normalized to its value at $\alpha=0^{\circ}$. As one can see, the power is strongly reduced with the increase of the angle between laser axes. It is explained by the sufficient decrease of the volume in which plasma wakes are overlapped. For the proof-of-principal experiment to be implemented,   the small angle ($\sim$ several degrees) between laser pulses is required. For the discussed parameters, the angle of $\sim5^{\circ}$ is appropriate to keep the balance between efficient enough radiation and the possibility to realize this idea experimentally. At $\alpha=5^{\circ}$, the radiation power decreases by $20$\% compared to the case of the counterpropagating laser pulses and reaches $\sim40$ MW. 

\subsection{Angular distribution of the radiation intensity}
\begin{figure*}[htb]
	\includegraphics[width=1.\linewidth]{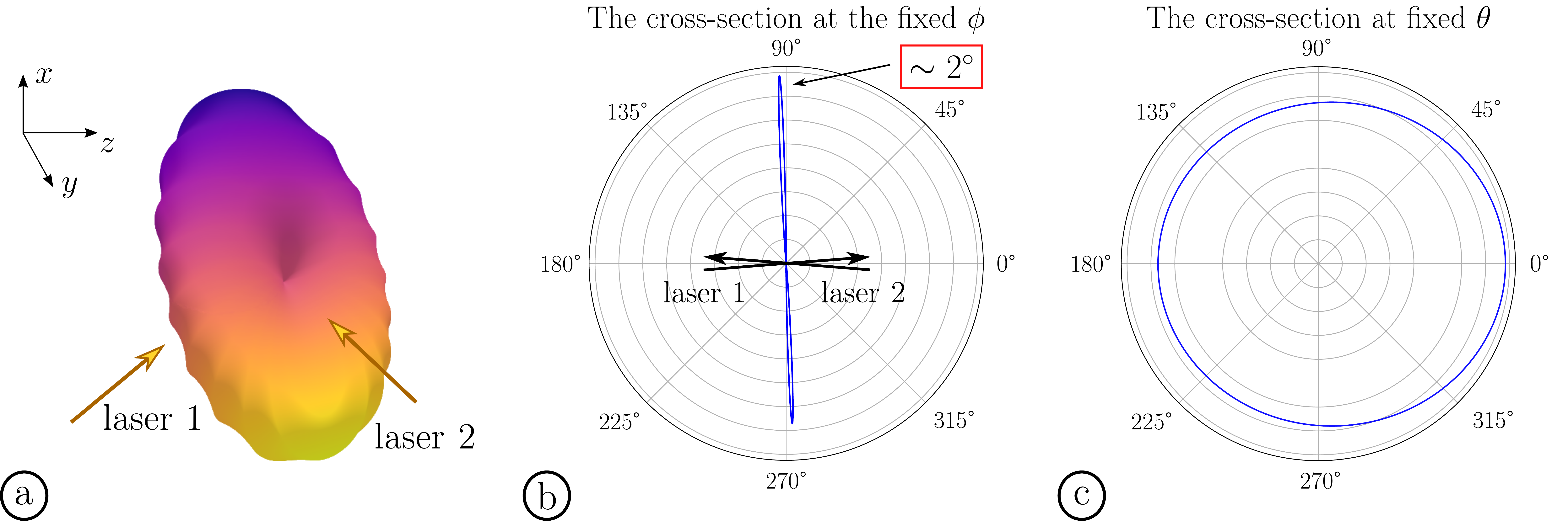}
	\caption{Angular distribution of radiation intensity at $\alpha=5^{\circ}$: a schematic 3D distribution (a); distribution over the polar angle in $(x,z)$-coordinates (b); distribution over the azimuthal angle in the frame co-directed to laser 1 axis (c).}\label{fig:Distr}
\end{figure*}

\begin{figure}[htb]
	\includegraphics[width=1.\linewidth]{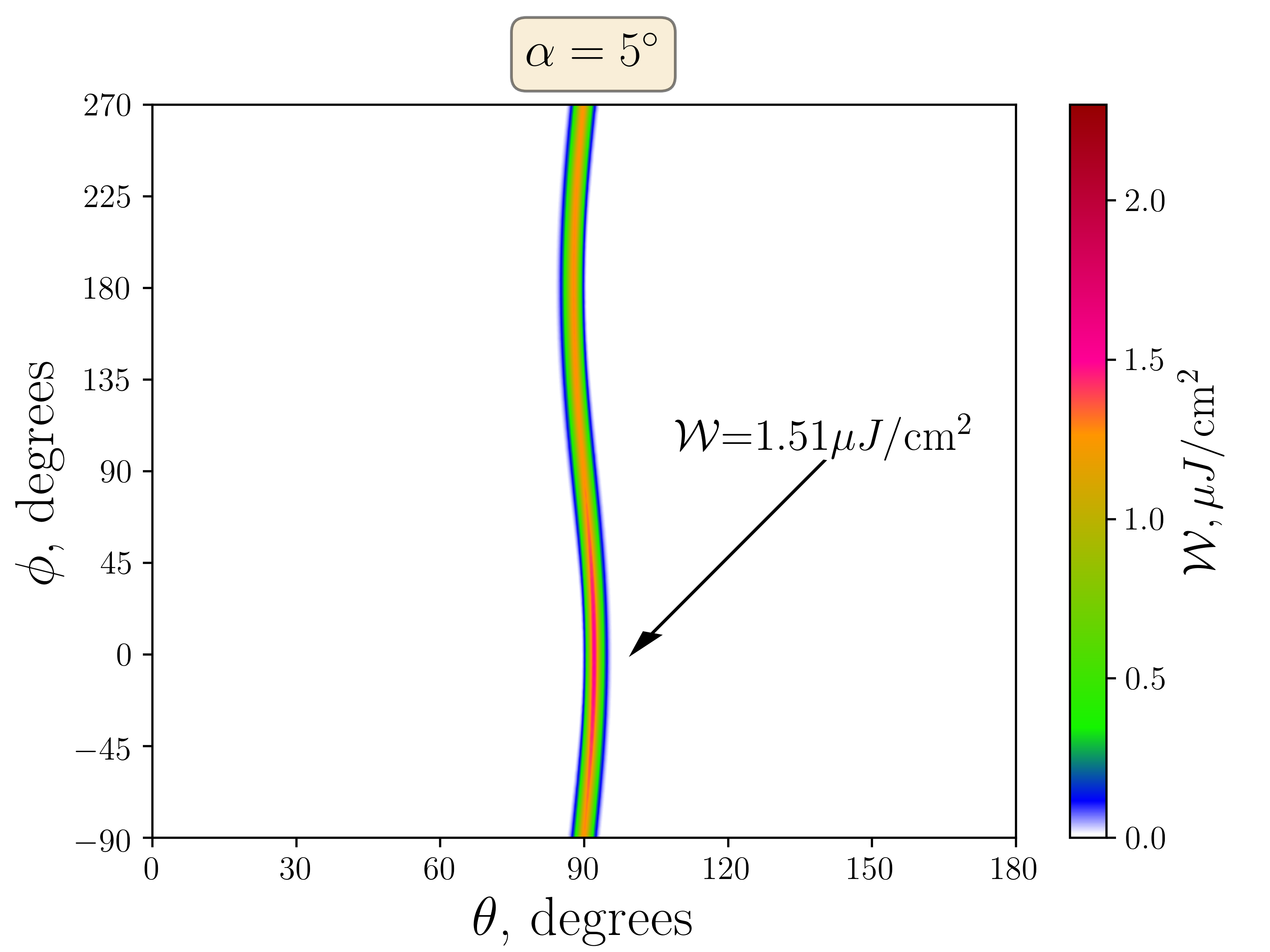}
	\caption{Map of radiation energy density $\mathcal{W}(\theta,\phi)$ at the distance of $R=10$ cm for the collision angle $5^{\circ}$.}
	\label{fig:DistrPl}
\end{figure}

The angular distribution of the radiation intensity for this case is presented in Figure \ref{fig:Distr}. In the far zone of the source, radiation pattern looks 
like a thin disc (Fig.\ref{fig:Distr} (a)) oriented transversely to the axis of the narrow laser pulse (laser 1). Being almost uniformly distributed over the 
azimuthal angle (Fig.\ref{fig:Distr} (c)), it is concentrated in a narrow cone of about $2^{\circ}$ over the polar $\theta$-angle (Fig.\ref{fig:Distr} (b)). Such a narrow distribution corresponds to the diffraction angle $\lambda/D\approx 1.8^{\circ}$ where $\lambda=\pi c/\omega_p$ is the radiation wavelength and $D\sim 100 c/\omega_p$ is the typical length of radiating plasma.  
As the angle between laser axes increases, the direction of the maximal intensity remains oriented transversely to the laser 1 axis. Thus, the direction of the most intense radiation deviates from $\theta=90^{\circ}$ by the angle $\beta$. The asymmetry of the system leads to the asymmetry of the radiation pattern: the disc is more extended towards the alignment of the laser axes, which is also seen from Figure \ref{fig:DistrPl}.  At small $\alpha$, this effect is seen to be weak.

\begin{figure*}[ht!]
	\includegraphics[width=1.\linewidth]{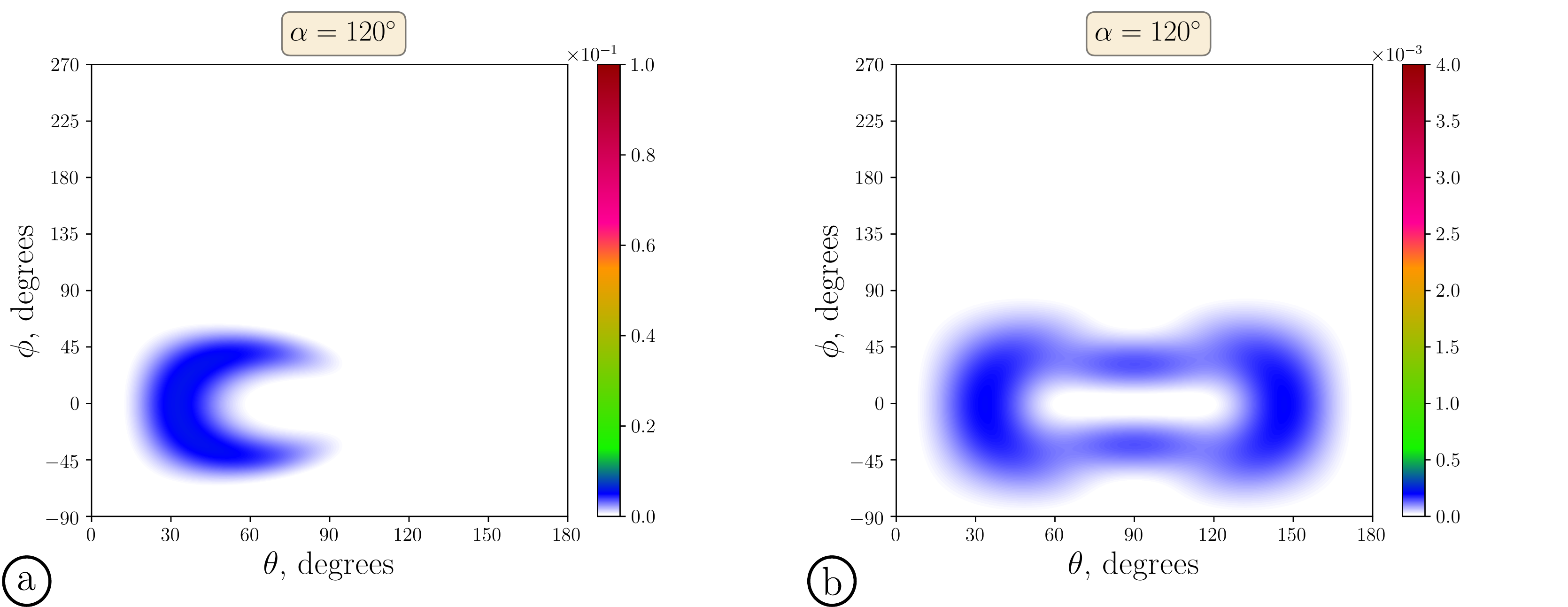}
	\caption{Angular distributions of radiation intensity in the far zone at the collision angle $\alpha=120^{\circ}$ for asymmetric (involved in the present work case) (a) and symmetric (b) laser pulses.}
	\label{fig:Distr120}
\end{figure*}
In order to choose the detector with the correct sensitivity for measuring $2\omega_p$-emission in the demonstration experiment, we should calculate the energy coming through the unit square at the detector location. Let us estimate this value at the distance of order of $10$ cm from the source where a diagnostic equipment will be placed. In the work \cite{timofeev2020simulations},  particle-in-cell simulations of a self-consistently evolving laser pulse have demonstrated that theoretical predictions overestimate the efficiency of the laser-to-radiation energy conversion by $30$ \%. We will assume that the theoretical description proposed here predicts the correct spatial distribution of radiation, but overestimates the absolute value of the radiated energy. According to this, we can estimate the total energy  radiated with EM waves into the whole sphere $4\pi$  as $\sim 40 \mu \mbox{J}$. The dependence of energy density distribution $\mathcal{W}$ on angles $\theta$ and $\phi$ at the collision angle $\alpha=5^{\circ}$ is presented in Figure \ref{fig:DistrPl}. This Figure clearly demonstrates the asymmetry of the distribution: the radiation is predominantly emitted towards the alignment of laser axes. Granted this energy distributed according our theoretical predictions, the maximal energy density at the distance 10 cm from the source reaches $\sim 1.5 \mu J/\mbox{cm}^2$.

\subsection{EM emission due to three-wave interaction}

It is interesting to note a slight increase in the radiation power at $\alpha=120^{\circ}$ (Fig. \ref{fig:Power}). We associate this increase with three-wave interaction of plasma waves. Indeed, if we replace laser-driven wakefields  characterized by some nonuniform transverse potential profiles with plane Langmuir waves $(\omega,{\bf k_1})$ and $(\omega,{\bf k_2})$, three-wave synchronism conditions with EM wave $(2\omega, {\bf k_1}+{\bf k_2})$ become met:
\begin{equation}
	\omega+\omega=\sqrt{1+k_{EM}^2}, \qquad k_{EM}=|\textbf{k}_1+\textbf{k}_2|=\sqrt{3}.
\end{equation}
Despite of falling in resonance, such an EM wave is not generated, since the nonlinear electric current produced by electrostatic oscillations is directed along the axis of EM wave propagation and therefore cannot do work on its transversely polarized electric field. That is the reason why we do not observe radiation near the point $(\theta =\pi /2, \phi = 0)$ in Figure \ref{fig:Distr120} (a) that shows the angular distribution of $2\omega_p$ emission calculated in our theory for $\alpha=120^{\circ}$.  In reality,  the influence of the transverse structure of real wakes  and the spatial limitation of the radiation region lead to broadening of the three-wave resonance. Therefore, generation of EM waves deviating from the direction $(\theta =\pi /2, \phi = 0)$ becomes possible. This mechanism is seen not to contribute significantly to the total radiation power. The asymmetry of the distribution in this case is associated with the asymmetry of the laser pulses. For equal focal spots of colliding pulses, the angular distribution turns out to be symmetric, although its intensity is two orders of magnitude lower  (Figure \ref{fig:Distr120} (b)).

\section{Numerical simulations \label{Sec:Simulations}}

\begin{figure}[h]
	\includegraphics[width=0.99\linewidth]{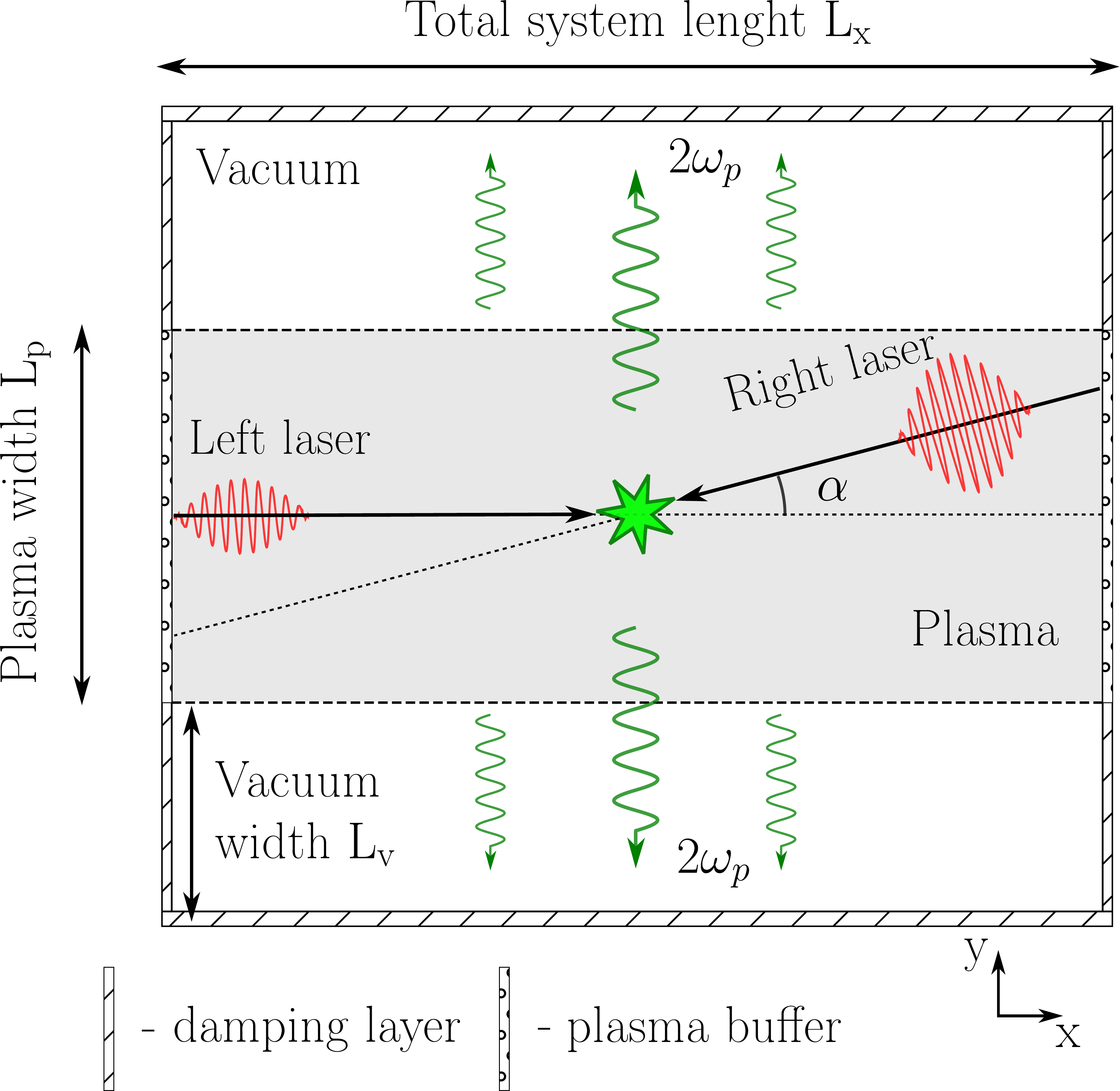}
	\caption{Simulation layout. A narrow laser pulse (left) propagates along the axis of the system while a wider pulse (right) is oriented at an angle $\alpha$ to the first one. Both laser pulses are focused in the system center. The produced radiated is absorbed in damping layers at the top and bottom of the system.   }
	\label{fig:layout}
\end{figure}

In the head-on collision case, theoretical predictions have been verified by particle-in-cell simulations in plane (2D Cartesian) geometry \cite{timofeev2017generation}. In the paper \cite{timofeev2020simulations}, the same problem has been studied numerically for an axially symmetric system. Simulations in both geometries have demonstrated good agreement with our theoretical insights. In order to verify the theory presented in the current work, it is required to realize full 3D geometry. Such simulations are rather resource-intensive and are not available for us at this moment due to a lack of computing resources. For this reason, in this section, we simulate the oblique collision of plasma wakes in plane geometry. The approach involved does not allow complete quantitative comparison, but it is suitable for qualitative analysis of the certain conclusions of the theory. In particular, we are interested in verifying our theoretical predictions about the direction of EM emission. In addition, results obtained in plane geometry can be applied for the collision of laser pulses focused to the strongly stretched in one direction rectangular spots. This possibility has been also discussed in the work \cite{timofeev2017generation}.
\begin{table}[h!]
	\centering
	\caption{Parameters of simulations.}
	\begin{ruledtabular}
		\begin{tabular}{lcc}\label{Table:PICParams}
			
			Grid size & $\Delta x=\Delta y$	&  $0.01 c/\omega_p$\\
			
			Time step & $\Delta t$	&  $0.005 \omega_p^{-1}$\\
			
			Total system length &$L_x$ & $4800$ cells\\
			
			Plasma width & $L_p$ & $2500$ cells\\
			
			Vacuum layers width & $L_v$ & $1502$ cells\\
			
			Damping layers width & $L_d$ & $100$ cells\\
			
		\end{tabular}	
	\end{ruledtabular}
\end{table}
To simulate generation of $2\omega_p$-radiation by colliding wakes, we use a standard 2D3V particle-in-cell code. Its description can be found, for example, in the paper \cite{annenkov2018high}. The layout of the simulation box is shown in Figure \ref{fig:layout}. Laser pulses are injected in a plasma layer with a width $L_p$ towards each other. Since the radiation is expected to be oriented across a narrower laser pulse, this pulse propagates along the axis of the system while a wider pulse is oriented at an angle $\alpha$ to the first one. In this case, we expect that radiation will propagate strictly across the plasma and be absorbed in the top and bottom damping layers. Interaction of laser radiation with a plasma is described by the slowly varying ponderomotive force $\textbf{F}_s=-\nabla |a_s(\textbf{r},t)|^2/4$, both laser pulses are focused in the center of the simulation box. To decrease the influence of edge effects arising while a virtual laser pulse crosses over a sharp plasma boundary, the ponderomotive force of each pulse gradually increases for the first 300 cells. Ions are immobile with electrons being distributed by Maxwell distribution function with an initial temperature $T_e=14$ eV. Main parameters of the computational scheme are presented in Table \ref{Table:PICParams}. Since the system size is rather large,  main simulations were carried out with 9 particles per cells (ppc). To justify the advisability of such a choice, simulations with the same physical parameters were also performed with 25 and 49 ppc for the case of $\alpha=7.5^{\circ}$, as well as with a smaller grid $\Delta x=\Delta y=0.005, \Delta t=0.0025$. The difference in the radiation efficiency in all these runs  is found to be negligible, which justifies the use of 9 ppc.

\begin{figure}[htb]
	\includegraphics[width=0.8\linewidth]{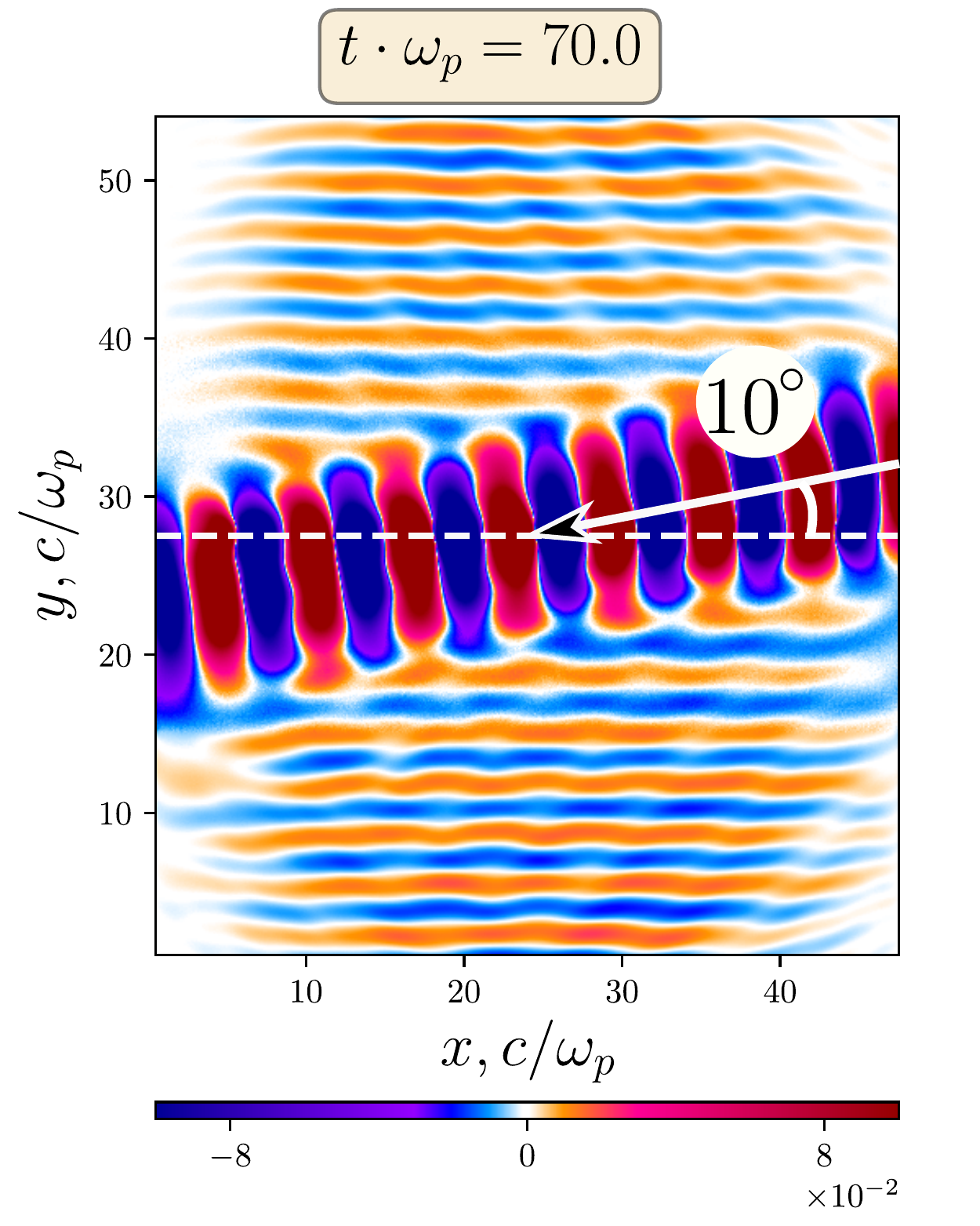}
	\caption{Electric field map $E_x(x,y)$ in PIC simulations at $\alpha=10^{\circ}$.}
	\label{fig:PICEx}
\end{figure}

\begin{figure*}[htb]
	\includegraphics[width=0.95\linewidth]{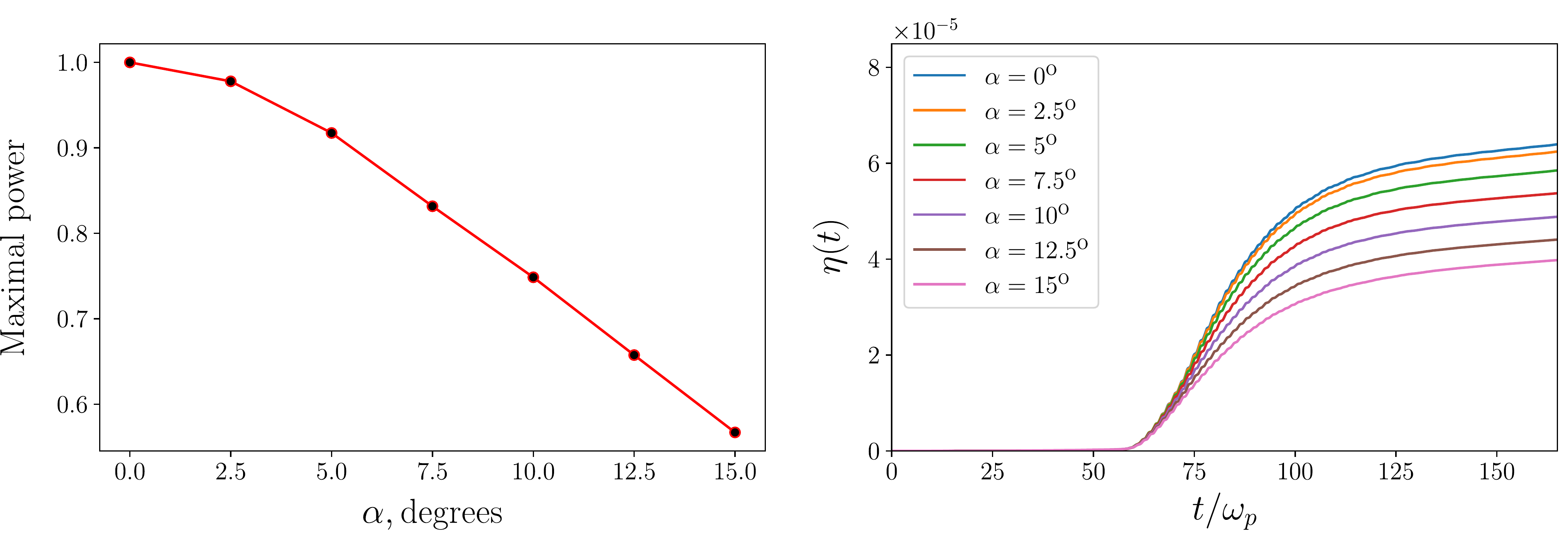}
	\caption{Left: the maximal radiation power in  units of the head-on collision power (at $\alpha=0^{\circ}$) as a function of collision angle; Right: the temporal dependence of the radiation efficiency at different angles.} 
	\label{fig:eta_a}
\end{figure*}

To demonstrate the qualitative agreement with the theory, a series of PIC simulations for different collision angles has been carried out. Given other equal conditions, the angle $\alpha$ varies in  the range from $0^{\circ}$ to $15^{\circ}$ with a step of $2.5^{\circ}$. Figure \ref{fig:PICEx} shows that the observed radiation is really oriented across the narrower laser pulse in agreement with the theory. 
In Figure \ref{fig:eta_a} (left), one can see how the maximal (over time) radiation power  normalized to the power at $\alpha=0^{\circ}$ depends on the collision angle. The hystory of the radiation efficiency being determined as the relation of energy absorbed in damping layers to the total laser energy is shown in Figure \ref{fig:eta_a} (right). With the increase of $\alpha$, the efficiency of radiation decreases but more slowly (by $\sim10$\% at $\alpha=5^{\circ}$) than was obtained in theory. This may be due to the peculiarities of plane geometry.

\section{Conclusion \label{Sec:Conclusion}}

In this paper, the theory of EM emission at the second harmonic of plasma frequency produced by colliding laser-driven plasma wakes is generalized to the case of an arbitrary collision angle $\alpha$. We calculated the angular distribution of radiation intensity in the far zone of the source for the parameters of the proof-of-principle experiment on the collision of different-size laser pulses with the total energy 0.2 J in a supersonic gas jet, which is started in the Institute of Laser Physics SB RAS. It is found that radiation pattern looks like a thin disc which, at small $\alpha$, has the very narrow distribution over the polar angle $2^{\circ}$. The direction of the maximal intensity is shown to be always oriented transversely to the axis of a narrower laser pulse. This theoretical result is confirmed by particle-in-cell simulations. The total radiation power rapidly decreases with the increase of the collision angle so that, at  $\alpha=5^{\circ}$, it is reduced by 20\%. Thus, the proposed theory not only answers the question where the produced second harmonic radiation should be detected in this experiment, but also formulates the requirements to both the angle of laser alignment ($<5^{\circ}$) and sensitivity of a detector (the maximal density of radiated energy at the distance 10 cm from the laser focus is estimated at the level 1.5 $\mu \mbox{J/cm}^2$).

\section{ACKNOWLEDGMENTS}
Simulations were performed using the computing resources of the Center for Scientific IT-services ICT SB RAS (https://sits.ict.sc). 
This work is supported by the Russian Foundation for Basic Research (grant 20-32-70055). 

\newpage
\appendix
\section{Radiation current \label{sec:appendix}}
The nonlinear current (\ref{current}) expressed in terms of scalar potentials of colliding plasma wakes can be represented in the following form:
\begin{widetext}
\begin{multline}
\mathcal{J}_x=\frac{e^{2i\sin\beta}}{4}\left[\sin\beta(1+\cos^2\beta)\left(\Phi_1\Phi_{2_{xx}}+\Phi_2\Phi_{1_{xx}}\right)+\sin\beta\left(\Phi_1\Phi_{2_{yy}}+\Phi_2\Phi_{1_{yy}}\right)+\sin\beta(3-4\sin^2\beta)\Phi_{1_{x}}\Phi_{2_{x}}+\right.\\
 \left.+\Phi_{1_{y}}\Phi_{2_{y}}\sin\beta-\sin\beta(1+2\sin^2\beta)\Phi_1\Phi_2-i\cos\beta\left(1+\frac{\cos \alpha}{2}\right)\left(\Phi_{1_{x}}\Phi_{2_{xx}}+\Phi_{2 _{x}}\Phi_{1_{xx}}\right)-i\cos\beta\left(\Phi_{1_{x}}\Phi_{2_{yy}}+\Phi_{2 _{x}}\Phi_{1_{yy}}\right)-\right.\\
 \left.-\frac{i}{2}\cos\beta\left(\Phi_{1_{y}}\Phi_{2_{xy}}+\Phi_{2 _{y}}\Phi_{1_{xy}}\right)+i\cos\beta\left(\frac{1}{2}+3\sin^2\beta\right)\left(\Phi_1\Phi_{2_{x}}+\Phi_2\Phi_{1_{x}}\right)\right], 
\end{multline}

\begin{multline}
\mathcal{J}_y=\frac{e^{2i\sin\beta}}{4}\left[\sin\beta\cos\beta\left(\Phi_2\Phi_{1_{xy}}+\Phi_1 \Phi_{2_{xy}}+\Phi_{1_{x}} \Phi_{2 _{y}}+\Phi_{1_{y}} \Phi_{2_{x}}\right)-i\left(\Phi_{1_{y}}\Phi_{2_{xx}}+\Phi_{2 _{y}}\Phi_{1_{xx}}\right)-\right.\\
\left.-\frac{3}{2}i\left(\Phi_{1_{y}}\Phi_{2_{yy}}+\Phi_{2 _{y}}\Phi_{1_{yy}}\right)-\frac{i}{2}\cos \alpha\left(\Phi_{1_{x}}\Phi_{2_{xy}}+\right.\left.+\Phi_{2_{x}}\Phi_{1_{xy}}\right)+i\left(1-\frac{\cos  \alpha}{2}\right)\left(\Phi_1 \Phi_{2_{y}}+\Phi_2\Phi_{1_{y}}\right)\right],
\end{multline}
\begin{multline}
\mathcal{J}_z=\frac{e^{2i\sin\beta}}{4}\left[\cos\beta(1+\sin^2\beta)\left(\Phi_1\Phi_{2_{xx}}-\Phi_2\Phi_{1_{xx}}\right)+\cos\beta\left(\Phi_1\Phi_{2_{yy}}-\Phi_2\Phi_{1_{yy}}\right)+i \sin\beta \left(1-\frac{\cos  \alpha}{2}\right)\left(\Phi_1\Phi_{2_{x}}-\right.\right.\\
\left.\left.-\Phi_2\Phi_{1_{x}}+\Phi_{1_{x}}\Phi_{2_{xx}}-\Phi_{2_{x}}\Phi_{1_{xx}}\right)-i\sin\beta\left(\Phi_{2_{x}}\Phi_{1_{yy}}-\Phi_{1_{x}}\Phi_{2_{yy}}\right)-\frac{i}{2}\cos\beta\left(\Phi_{1_{y}}\Phi_{2_{xy}}-\Phi_{2_{y}}\Phi_{1_{xy}}\right)\right],
\end{multline}
\end{widetext}
where $\Phi_{1_x}, \Phi_{2_y}$ and so on denote partial derivatives with respect to the corresponding coordinates, i.e, for example $\Phi_{1_x}=\dfrac{\partial \Phi_1}{\partial x}.$

\bibliography{bibLasers}

\end{document}